# Angular Trapping of Spherical Janus Particles

Xiaoqing Gao,[1,2] Yali Wang,[3] Xuehao He,[3] Mengjun Xu,[4] Jintao Zhu,[4] Xiaodong Hu[1], Xiaotang Hu[1], Hongbin Li[1,2]* and Chunguang Hu[1]*

[1]State Key Laboratory of Precision Measuring Technology and Instruments

School of Precision Instrument and Optoelectronics Engineering

Tianjin University

Tianjin, 300072, P. R. China

[2]Department of Chemistry

University of British Columbia

Vancouver, BC V6T 1Z1

Canada

[3]Department of Chemistry

School of Science

Tianjin University

Tianjin, 300072, P. R. China

[4]School of Chemistry and Chemical Engineering

Huazhong University of Science and Technology

Wuhan, 430074, P. R. China

*To whom correspondence should be addressed to C. H. (cghu@tju.edu.cn) or H. L. (hongbin@chem.ubc.ca).

**Abstract**

Developing angular trapping methods, which will enable optical tweezers to rotate a micronsized bead, is of great importance for the studies of biomacromolecules during a wide range of torque-generation processes. Here we report a novel controlled angular trapping method based on composite Janus particles. We used a chemically synthesized Janus particle, which consists of two hemispheres made of polystyrene (PS) and poly(methyl methacrylate) (PMMA) respectively, as a model system to demonstrate this method. Through computational and experimental studies, we demonstrated the feasibility to control the rotation of a Janus particle in a linearly polarized laser trap. Our results showed that the Janus particle aligned its two hemisphere's interface parallel to the laser propagation direction as well as the laser polarization direction. In our experiments, the rotational state of the particle can be easily and directly visualized by using a CMOS camera, and does not require complex optical detection system. The rotation of the Janus particle in the laser trap can be fully controlled in real time by controlling the laser polarization direction. Our newly developed angular trapping technique has the great advantage of easy implementation and real time controllability. Considering the easy chemical synthesis of Janus particles and implementation of the angular trapping, this novel method has the potential of becoming a general angular trapping method. We anticipate that this new method will significantly broaden the availability of angular trapping in the biophysics community, and expand the scope of the research that can be enabled by the angular trapping approach.

Optical tweezers have become an indispensable tool in single molecule manipulation, allowing for the investigation of the mechanics and conformational dynamics of biomacromolecules, including nucleic acids and proteins, at the single molecule level [1-7]. Most single molecule manipulations by optical tweezers are realized via three-dimensional translation of the trapped micrometer sized bead to which a single biomacromolecule is attached, and the resultant force experienced by the biomacromolecule can then be measured. It has long been recognized that entailing optical tweezers an ability to rotate a biomacromolecule will be equally useful [8-12]. Many attempts have been made to realize angular trapping of particles [13,14]. Amongst these efforts, angular trapping has been successfully implemented using birefringent particles, including calcite fragments [8], vaterite microspheres [10,15-19] and fabricated quartz cylinders [9,20-27], leading to the controlled rotation of such particles in optical tweezers. Due to its chemical stability and easy surface functionalization, nanofabricated quartz cylinders-based angular trapping has been successfully used in single molecule biophysics studies [9,12,21-26]. Despite its success, the use of quartz cylinders-based angular trapping remains limited, largely due to the challenges in the nanofabrication of quartz cylinders as well as the low birefringence of quartz, which leads to a low transfer angular momentum efficiency [27]. To expand the application of angular trapping to a broader community, developing new angular trapping methodologies is of critical importance.[27] Here we demonstrate a novel angular trapping method using chemically synthesized spherical Janus particles. This novel method can be easily implemented and the rotation of the Janus particle can be directly visualized and measured using a complementary metal–oxide–semiconductor (CMOS) camera. Given the ease of the chemical synthesis of the Janus particles as well as simple implementation of the angular trapping, we anticipate that this new method will find a wide range of applications in single molecule biophysical studies.

Significant advance in chemical synthesis of Janus particles has been achieved over the last decade[28]. It has become possible to synthesize Janus particles with exquisite control of composition, shape and surface properties. Janus particles are special types of nano/microparticles whose surfaces show two or more distinct physical and chemical properties. The simplest Janus particle is a spherical particle that consists of two chemically/physically distinct hemispheres, such as refractive index.[29] Such Janus particles can be considered as birefringent. Realizing that birefringent particles could be aligned by a linear polarized laser light, we sought to develop a general angular trapping strategy based on such Janus particles.

The Janus composite particles we used here were chemically synthesized via a microcapillary-based microfluidic method (Fig. S1)[29]. They are composed of two hemispheres made of polystyrene (PS) with a refractive index of 1.57 and poly(methyl methacrylate) (PMMA) with a refractive index of 1.48, respectively. The diameter of the Janus composite particles was ~4 μm. Fig. 1 shows the optical image of this Janus composite particle. Due to the difference in their refractive index, the PS hemisphere with a higher refractive index appeared darker than the PMMA hemisphere with a lower refractive index. The interface between the PS and PMMA hemispheres can be directly visualized under an optical microscope, providing a convenient way to visualize the particle rotational state in angular trapping experiments.

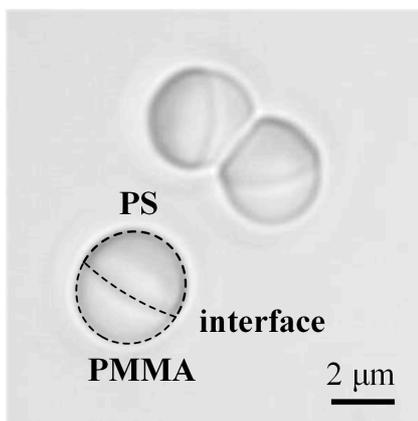

Figure 1. Optical images of the PMMA/PS Janus particle. The diameter of the particle is ~4 μm. The PS hemisphere has a higher refractive index and appears darker. The lighter hemisphere corresponds to PMMA.

To investigate the feasibility of controlling such Janus particles in a linearly polarized optical tweezers, we first examined the trapping of the Janus particle computationally. Although optical force has long been used to understand the optical trapping (including angular trapping)[30-38], such analysis proved challenging in dealing with Janus particles with inhomogeneous optical properties. Due to the distinct interface of the two hemispheres with different refractive index, multiple reflection and refraction will occur in the Janus particle, making quantitative calculations based on optical force very complex. To overcome this difficulty, here we exploited the energy minimization approach to quantitatively predict the trapping of a Janus particle in linearly polarized optical tweezers.

It is well recognized that the stable trapping position and orientation of a micronsized bead is achieved by maximizing the overlapping region of the optical field and particle[37,39]. The principle underlying the stable trapping state of a trapped particle is the minimization of the electromagnetic potential energy, which is related not only to the volume of the overlapping region, but also to the refractive index and optical intensity distribution[37,39]

In an electromagnetic field, the electromagnetic energy of the optical field is given by the volume integral of $\mathbf{D} \cdot \mathbf{E}$, where $\mathbf{D}$ is the electric displacement and $\mathbf{E}$ is the electric field. When a dielectric particle with permittivity of $\varepsilon_p$ is present in the optical field in a medium with permittivity of $\varepsilon_m$, the energy of the occupied region is reduced to $\varepsilon_m/\varepsilon_p$ of the original energy. For a non-magnetic particle, the permittivity equals to the square of refractive index $n$, so the electromagnetic energy stored in the whole system is given as:

$$\langle U \rangle = \frac{1}{2}\varepsilon_0 n_m^2 \int E_m^2 \, dV_t - \frac{1}{2}\varepsilon_0 n_m^2 \int \left(1 - \frac{n_m^2}{n_p^2}\right) E_m^2 \, dV_p = U_t - \Delta U_p$$

where the angle brackets indicate time average; $\varepsilon_0$ is permittivity of the vacuum, $n_m$ and $n_p$ are refractive indexes of the medium and particle, respectively. $E_m$ is the electric field distribution in the absence of a particle, $V_t$ is the total volume of whole system and $V_p$ is the volume occupied by the trapped particle.

The energy equation contains two terms: the first term, which integrates over the whole optical field, represents the total energy $U_t$ stored in the system when no particle is present in the field of interest. The second term is the energy reduction $\Delta U_p$ due to the particle with a higher refractive index occupying the optical field. $\Delta U_p$ is determined by the electric field and refractive index of the overlapping region. For simplicity of the calculation, the energy for the whole optical trap field without any particle $U_t$ is defined as zero.

To calculate the electromagnetic energy $\langle U \rangle$, a coordination system was built around the trap (Fig. 2). The trap center is fixed at the origin and the laser propagates along the z axis. The focused laser field was described using the classical method by Wolf [40,41]. The trapping laser is a linearly polarized Gaussian beam with a wavelength of 1064 nm and the polarization direction is parallel to the x axis. The objective which focuses the trapping laser had a numerical aperture of 1.2. The whole system was immersed in water (with a refractive index of 1.33).

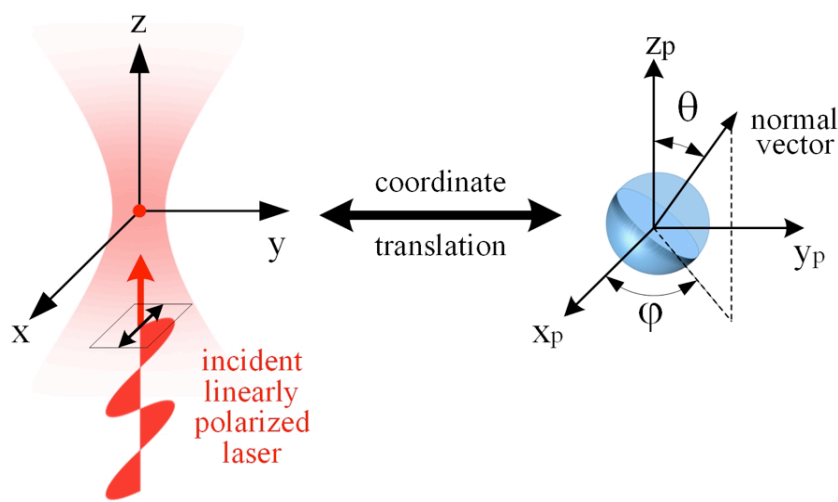

Figure 2. The coordinate system used in the computational study. The origin of the lab coordinate (x, y, z) is set at the trap center. The laser beam propagates along the z axis, and the polarization direction is along the x-axis. The particle center is defined by its coordinate (x, y, z). To define the rotational state of the Janus particle, the particle coordinate system ($x_p$, $y_p$, $z_p$) is also built, which is located in the center of the Janus particle and can be transformed to the lab coordinate by translation. In the particle coordinate, the two rotational DOFs of the Janus particle is described by the elevation angle $\theta$ and azimuth angle $\varphi$ of the virtual normal vector, which is perpendicular to the two hemisphere interface and located in the center of the Janus particle.

Due to its rotational symmetry, the Janus particle has five degrees of freedoms (DOFs). The three translational DOFs are described by the coordinate of the particle center (x, y, z) relative to the lab coordinate (Fig. 2). To describe the two rotational DOFs of the Janus particle, we also defined the particle coordinate ($x_p$, $y_p$ and $z_p$), which is located in the center of the Janus particle and can be transformed to the lab coordinate by translation. In the particle coordinate, the two rotational DOFs can then be described by the elevation angle θ and azimuth angle φ of the virtual normal vector. The virtual normal vector, which is located in the center of the particle, is defined as being perpendicular to the two hemisphere interface and pointing to the PMMA hemisphere with a lower refractive index (Fig. 2).

Having established the coordinate, we first tested this energy minimization approach for the trapping of a homogenous spherical particle, i. e. $n_1=n_2$. Given the three plane symmetry of the optical field (with respect to xy, xz and yz planes) (Fig. S2), it can be readily shown that the potential energy ⟨U⟩ reaches its minimum only when the homogeneous particle is trapped in the focal point of the laser beam. In other words, the geometric center of the bead (x, y, z) superimposes with the focal point of the laser trap (Fig. S3). This result confirmed the validity of this approach for the trapping of a homogeneous spherical particle.

We then applied this approach to calculate the total potential energy ⟨U⟩ of the optical field with the Janus particle. For the linearly polarized laser beam used in our optical tweezers, its intensity profile is not the same in the x, y and z direction, rather, it extends the most in the direction of laser propagation (z-direction), then the polarization direction (x-direction) and last in the y directions (Fig. S2). Considering that the optical field is symmetric with respect to the xy plane (when z=0) and the stable trapping position of the Janus particle in the optical trap should be close to the trapping center, it becomes evident that in order to maximize the overlap with the optical field, the Janus particle must be located in the focal plane in its stable trapping position, i.e. z=0. To confirm this reasoning, we calculated the electromagnetic energy ⟨U⟩ of the optical field with a Janus particle placed at yz plane (x=0) and xz plane (y=0), respectively. At each plane, we calculated ⟨U⟩ at each coordinate by exhausting the possible choices of the other four DOFs, and then determined the ⟨U⟩$_{min}$ at each coordinate (x, z) in the xz plane or (y, z) in the yz plane. Fig. 3A-3B plots ⟨U⟩$_{min}$ at each coordinate in the xz and yz plane. It is evident that indeed the electromagnetic energy ⟨U⟩ achieved the local minima when z=0 in both planes.

Based on this feature, we then calculated the potential energy ⟨U⟩ of the optical field with a trapped Janus particle placed at z=0 (i.e. in the xy plane) at each coordinate by exhausting the choices of the other four DOFs, and determined the minima at each coordinate (Fig. 3C). Our calculations predicted two energy minima with the same energy, i.e. two stable trapping states of the Janus particle. At these two stable states, the particle was located at the (x, y, z, θ, φ) coordinates of (0, 0.45, 0, 90°, 90°) and (0, -0.45, 0, 90°, 270°) At these two coordinates, the trapped Janus particle aligned its interface with the plane defined by the laser propagation direction (z-direction) (θ= 90°) and the laser polarization direction (x-direction) (φ= 90° or 270°) (Fig. 3D-3E). It is also of note that at the stable trapping positions, x=0, z=0, but y≠0, indicating that at the stable trapping positions, the particle center did not overlap with the laser focal point, instead it was always located in the PS hemisphere with a higher refractive index.

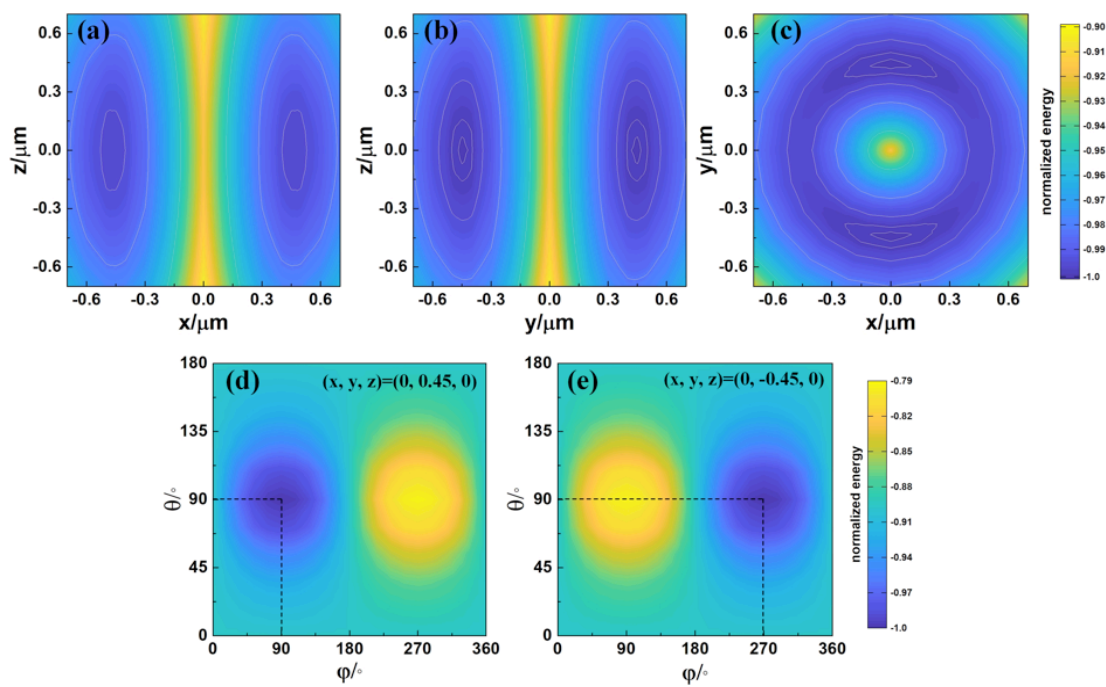

Figure 3. A-C) The minimum electromagnetic energy of the system with the Janus particle set in the xz plane (y=0) (A), yz plane (x=0) (B) and xy plane (z=0) (C). At each coordinate, the four DOFs were exhausted and the minimum energy is plotted in the figure. D-E) The profile of the minimum electromagnetic energy of the system with the Janus particle at xyz coordinate of (0, 0.45, 0) (D) and (0, -0.45, 0) (E).

The reason why the Janus particle showed specific stable orientations in the linearly polarized Gaussian beam can now be understood easily based on the energy minimization. Since the intensity of the optical field of a linearly polarized laser beam extends the most in the propagation (z-) direction, followed by polarization (x-) direction and last in the y-direction (Fig. S2), aligning the Janus particle with its interface parallel to the xz plane and the trapping center in the higher refractive index hemisphere would lead to the highest reduction of the electromagnetic energy. This is the result of a compromise between maximizing the overlapping volume of the particle with the laser beam and maximizing the overlap of high refractive index part with the optical field. Offsetting the trapping center from the geometric center of the particle leads to a slight reduction of the overall overlapping volume of the particle in the laser trap, but the offset towards the higher refractive index hemisphere increases the overlapping volume of the higher refractive index hemisphere with the laser field.

These results predicted that in the linearly polarized laser trap, the Janus particles will align itself to reach one of its two stable trapping positions, which were of the same energy. At these two stable trapping positions, the two hemisphere interface of the Janus particle is parallel to the plane defined by the laser propagation direction and polarization direction. Once the particle tilted away from the stable orientations, a restoring torque will be generated, trying to rotate the particle back to its stable position. The torque about the beam axis is induced by the optical intensity distribution due to laser polarization, and is the negative energy gradient along azimuth angle φ. If particle's interface deflects from laser polarization with an angle β, the restoring torque about beam axis is $A\sin(2\beta)$ (Fig. S4), where A is a constant and relates to trap intensity distribution, particle size and refractive index distribution.

Moreover, our calculations also predicted the response of the Janus particle in response to the change of the polarization direction. If the polarization direction is rotated away counterclockwise from the xz plane by an angle of α, our calculations showed that the new stable state will be located at coordinates of (-0.45sinα, 0.45cosα, 0, 90°, 90°+α) and (0.45sinα, -0.45cosα, 0, 90°, 270°+α). To reach these new stable positions, the Janus particle will rotate around the beam axis (z-axis) counterclockwise by an angle of α to align its particle interface with the direction of the new polarization direction. Since the particle center does not overlap with the trap center, such a rotation also resulted in a counterclockwise rotation of the particle center around the beam axis (z) by an angle of α. (Fig. S5). Rotating the polarization direction for 360° will lead to the rotation of the particle center by 360° following a circular trajectory (Fig. S5).

Taken together, our calculations strongly indicate that the Janus particle will take well-defined stable trapping positions and orientations in a linearly polarized laser trap, thus demonstrating the feasibility of angularly trapping a Janus particle in linearly polarized optical tweezers. To experimentally verify these predictions, we used the PS/PMMA composite Janus particle as the model system. The experiments were conducted on an optical tweezers system that has the same experimental parameters as the calculation. Fig. 4 shows the schematics of the angular trapping apparatus. The whole system is based on a standard single-beam optical trap plus a computer controlled half wave plate, which is placed before the objective to control the polarization of the laser beam. The polarization can be kept at a specific direction, or rotated at an angular velocity of up to 50 °/s. The position and orientation of the Janus particle were recorded directly using a CMOS camera in real time.

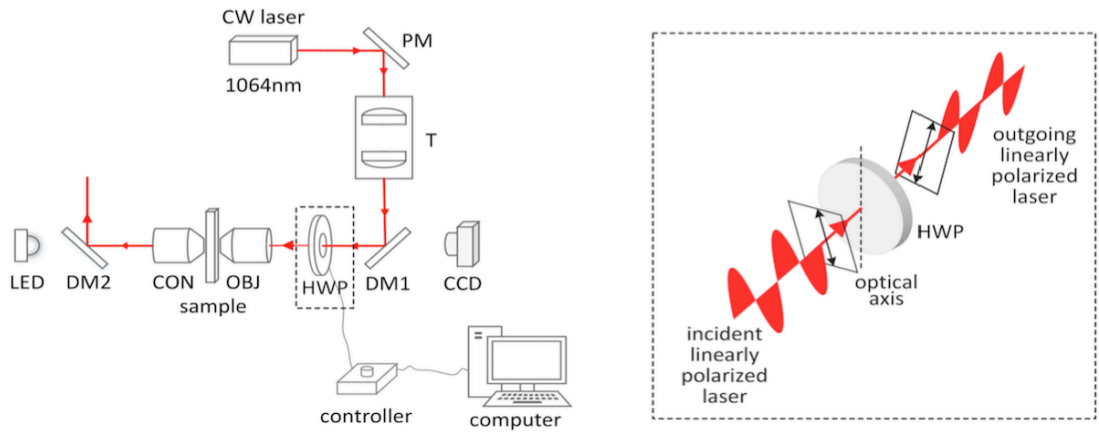

Figure 4. Schematics of the angular trapping setup. The laser trap is formed by a linearly polarized laser light (1064 nm). The laser polarization direction is controlled by a computer-controlled half-wave plate. The inset shows the schematics of controlling the laser polarization direction by the half-wave plate.

Just as predicted by our calculations, we observed that in the optical trap, the Janus PS/PMMA particle always aligned its interface parallel to the plane defined by the laser propagation and polarization direction in the linearly polarized optical trap (Fig. 5A). It is also interesting to observe that under each polarization direction, there were two distinct stable trapped orientations of the Janus particle at their stable trapping states. And these two orientations can overlap with each other by rotating 180 ° about the laser beam axis. Moreover, the choice of the two orientations by the Janus particle was random, and occurred roughly at a ratio of 1:1. The orientation of and the particle orientation. At each polarization direction, the Janus particle showed two stable orientations with the hemisphere interface always parallel to the laser polarization and propagation direction. B) The relationship of the interface angle and polarization angle. The orientation angle was obtained by processing 100 images of the Janus particle under a given polarization angle. The averages of the interface angles were plotted. The standard deviation of the measured interface angle was around 1-2°. C) Controlled angular trapping of the Janus particle by stepwise rotation of the polarization direction. D) Controlled angular trapping of the Janus particle by continuous rotation of the laser polarization direction in two different waveforms of the command. E) The trajectories of the geometrical center of the Janus particle during the change of the polarization direction by 360°. The trap center is located at the xy coordinate of (0, 0). Symbols are experimental data and the solid line is a fit to a circle.

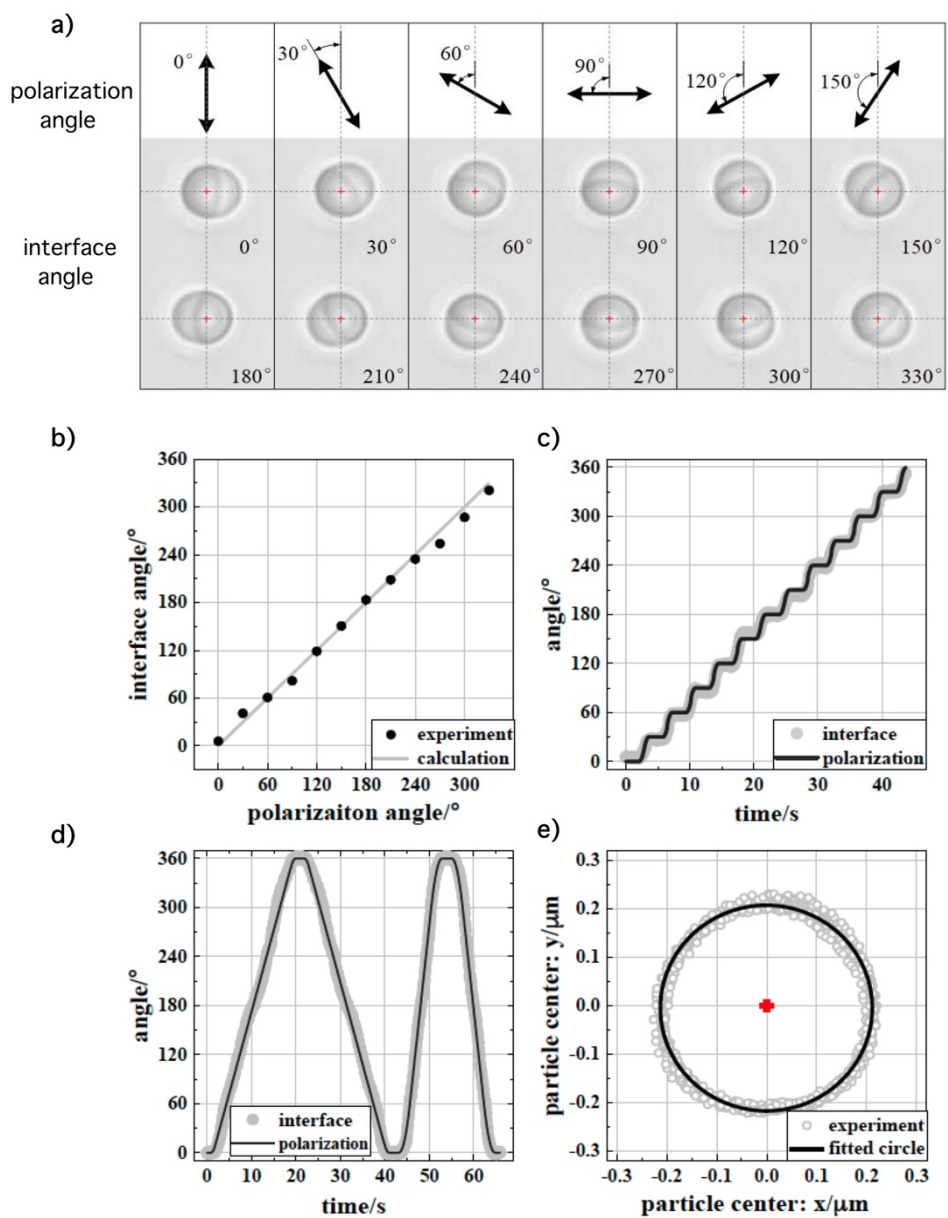

Figure 5. Angular trapping of the PMMA/PS Janus particle in a linearly polarized laser trap. A) Relationship of the laser polarization direction

the Janus particle can be controlled by external manipulation of the linear polarization direction of the laser bean via rotating the half wave plate (Fig. 5A and 5B). These results strongly indicated that there existed two stable trapped states of the Janus particle, the same as the predictions made by our calculations.

It is important to point out that the detection of the rotation of the Janus particles in our experiments was done by direct visualization in real time via a CMOS camera. Compared with the angular trapping with quartz cylinders[9,20,42], in which the rotation of the quartz cylinders has to be detected by measuring the change of the polarization direction, our angular trapping method has a much easier and simplified scheme to detect and measure the rotational state of the particle.

Having demonstrated the alignment of the trapped Janus particle in the linearly polarized laser trap, we then tested the controlled rotation of the Janus particle. The controllability we tested included two aspects: 1) to start or stop the rotation on demand; and 2) to control the rotational direction and angular velocity in real time. Fig. 5C-5D shows the results of these experiments. Clearly, the interface angle of the Janus particle was observed to closely follow the command of the polarization direction in real time.

Moreover, during the rotation of the Janus particle, the geometric center of the Janus particle was not stationary. Instead, it rotated around the beam axis. During a full 360º rotation of the Janus particle, the particle center followed a circular trajectory around the trap center with a radius of 0.20 μm, very close to the predicted value of 0.45 μm (Fig. 3D and Fig. S5). Our calculations indicated that the diameter of the circle is related to the particle radius and refractive index distribution.

In conclusion, using chemically synthesized Janus particle, we demonstrated a novel controlled angular trapping method both computationally and experimentally. Our results showed that the Janus particle aligned its interface parallel to the laser propagation direction as well as polarization direction in a linearly polarized laser trap. The rotational state of the particle can be easily and directly visualized by using a CMOS camera, and does not require complex optical detection system. The rotation of the Janus particle in the laser trap can be fully controlled in real time by rotating the laser polarization direction via rotating a half wave plate. Our calculations suggested that the achievable maximum torque about the beam axis is related to particle's refractive index distribution, implying that it is feasible to modulate the maximal generated torque by increasing the difference between the refractive index of the two hemispheres. This in turn can help modulate the temporal resolution of the angular trapping. Our newly developed angular trapping technique has the great advantage of easy implementation and real time controllability. Considering the relative easiness of the chemical synthesis of Janus particles and the possibility to tune their composition and physical properties, we anticipate that our angular trapping method will significantly broaden the availability of angular trapping in the biophysics community, and expand the scope of the research that can be enabled by this new angular trapping approach.

**Acknowledgements**

This work is supported by the National Natural Science Foundation of China. XG acknowledges the support of China Scholarship Council.